\documentclass[aps,pra,reprint, superscriptaddress]{revtex4-1}

\usepackage{epsfig,graphics,graphicx}
\usepackage[export]{adjustbox}
\usepackage{dcolumn}
\usepackage{bm}
\usepackage{color}
\usepackage{amsmath,amssymb,amsthm}
\usepackage{centernot}
\usepackage{bbold}
\usepackage{mathtools}
\usepackage{fullpage}
\usepackage{units}
\usepackage[usenames,dvipsnames]{xcolor}
\usepackage{braket}
\usepackage{indentfirst}

\usepackage[utf8]{inputenc}
\usepackage[T1]{fontenc}

\usepackage{mathpazo}
\usepackage{dsfont}

\usepackage{mathptmx}
\usepackage{amssymb}
\usepackage{amsmath}
\usepackage{latexsym}
\usepackage{amsfonts}

\usepackage{hyperref}
\hypersetup{pdfpagemode=UseNone}

\newtheorem{theorem}{Theorem}
\newtheorem{definition}[theorem]{Definition}

\newtheorem{lemma}[theorem]{Lemma}
\newtheorem{proposition}[theorem]{Proposition}

\newcounter{rem}
\newcommand{\mc}[1]{\mathcal{#1}}
\newcommand{\g}[1]{\mathfrak{#1}}

\def\>{\rangle}
\def\<{\langle}
\newcommand{\proj}[1]{| #1 \rangle\! \langle #1 |}

\renewcommand{\rho}{\varrho}
\DeclareMathOperator{\idty}{\mathbb{1}}

\def\ii{{\rm i}}

\DeclareMathOperator{\vspan}{span}
\def\textbf#1{{\bf #1}}
\newcommand{\Cx}{\mathbb{C}}
\newcommand{\Ir}{\mathbb{Z}}

\def\beq{\begin{equation}}
\def\eeq{\end{equation}}
\def\beqa{\begin{eqnarray}}
\def\eeqa{\end{eqnarray}}
\def\eea{\end{array}}
\def\bea{\begin{array}}
\newcommand{\bei}{\begin{itemize}}
	\newcommand{\eei}{\end{itemize}}
\newcommand{\bee}{\begin{enumerate}}
	\newcommand{\eee}{\end{enumerate}}
\def\bep{\begin{proposition}}
	\def\eep{\end{proposition}}
\def\bel{\begin{lemma}}
	\def\eel{\end{lemma}}
\def\bet{\begin{theorem}}
	\def\eet{\end{theorem}}
\def\bed{\begin{definition}}
	\def\eed{\end{definition}}

\usepackage{soul}

\begin{document}

\title{Quantum Walks via Quantum Cellular Automata}

\author{Pedro C.S. Costa}

\affiliation{Brazilian Center for Research in Physics-CBPF. Rua Dr. Xavier Sigaud, 150 - Urca - Rio de Janeiro - RJ - Brasil}

\author{Renato Portugal}
\affiliation{National Laboratory of Scientific Computing-LNCC. Av. Getúlio Vargas, 333 - Quitandinha, Petrópolis - RJ- Brasil}
\author{Fernando de Melo}
\affiliation{Brazilian Center for Research in Physics-CBPF. Rua Dr. Xavier Sigaud, 150 - Urca - Rio de Janeiro - RJ - Brasil}

\date{\today}


\begin{abstract}
Very much as its classical counterpart, quantum cellular automata are expected to be a great tool for simulating complex quantum systems. Here we introduce a partitioned model of quantum cellular automata and show how it can simulate, with the same amount of resources (in terms of effective Hilbert space dimension), various models of quantum walks. All the algorithms developed within  quantum walk models are thus directly inherited by the quantum cellular automata. The latter, however, has its structure based on local interactions between qubits, and as such it can be more suitable for present (and future) experimental implementations.
\end{abstract}


\maketitle
%


\section{Introduction}
Random walks, a certain type of stochastic processes, were introduced at the beginning of the twentieth century~\cite{RW}. Since then, random walks have turned into a powerful tool and found applications in areas as diverse as economics~\cite{Rw_economics}, ecology~\cite{RW_eco_paper}, computer science~\cite{RW_computer}, and physics~\cite{RW_phys1,RW_phys2}. Quantum mechanics was developed roughly at the same time. However, it took almost 90 years for the quantum version of random walks, nowadays known as \emph{quantum walks} (QWs) to be proposed~\cite{Coined}. Despite of its relative young age, the quantum counterpart of random walks is also being employed in various applications, most notably in quantum simulations~\cite{neutrino,curved} and in quantum algorithms~\cite{QWsearch}. Quantum walks, like its predecessor, comes in various flavors -- coined~\cite{Coined}, Szegedy~\cite{Szegedy1}, staggered~\cite{staggered}, and staggered with Hamiltonians~\cite{SQW_H}, to cite a few -- each with its own specificity and tuned to better deal with a given problem at hand. Foremost, some models of quantum walks were shown to form a universal platform for quantum computation~\cite{ChildsQW1,ChildsQW2}.
All this versatility and possible applications prompted an intense experimental activity with various realizations of quantum walks, for instance trapped ions~\cite{Trapped_ions}, optical lattices~\cite{QW_opitical} and more ~\cite{QW_implementations}. 

Most of quantum walk models, however, suffer from two major implementation drawbacks: \emph{i)} the space associated to the walkers  position is usually large, and as such it requires a high dimensional quantum system to encode it. \emph{ii)} The unitary evolving the walker is a global one, being applied in all the space in every step. These two points together tremendously limit the size of experimental implementations, and fundamentally  put the promising advantages of quantum walks into the far future.

On the classical world, the computational model of \emph{Cellular Automata (CA)} is also largely employed~\cite{CAbook}. Cellular automata are abstract dynamical/computational systems with discrete space and time. The space  consists of a $d$-dimensional network of cells. At each time step individual cells assume one state from a finite number of possible  states. The evolution of the network is given in parallel at discrete time steps: at each time $t$, the state of each cell depends only on the state of its neighboring cells at time $t-1$ through a local update rule~\cite{Wolfran}. Classical cellular automata, first proposed by John von Neumann and Stanislaw Ulam~\cite{Neumann}, are used in applications ranging from cryptography~\cite{CA_cripto} up to the simulation of fluid dynamics~\cite{FHP}, and the description of biological systems~\cite{CA_bio}.

Despite of all the results for classical cellular automata, its quantum counterpart, the quantum cellular automata (QCA), does not share the same amount of research activity. The phenomenology and applications of quantum cellular automata are very scarce. One possible reason for the QCA to not be yet widely employed is that a clear definition, that inherits the main qualities of classical CAs while allowing for a more powerful quantum processing, was not available until recently. The idea of quantum cellular automata was first pursued by Gröss and Zeilinger~\cite{QCAZeilinger}, with a model that later proved to not abide by the locality constraint expected of cellular automata~\cite{QCAmeyer}. After this first tentative various models of quantum cellular automata were put forward~\cite{QCAreview}. After a period of discussion about the proper quantization of cellular automata, in recent years an agreement was obtained and many of the definitions were proved equivalent~\cite{QCA_equivalent}.

The main goal of this contribution is to start exploring the model of quantum cellular automata, as a suitable platform for quantum algorithms. We do so by showing that various models of quantum walks can be translated to a quantum cellular automata dynamics using the same amount of resources (effective Hilbert space dimension). More concretely we show how to implement two models of QWs -- namely, the coined and the staggered QW with Hamiltonians, which are known to encompass a large class of QW models~\cite{SQW_H} -- within the partitioned QCA model (defined below). Quantum cellular automata, besides employing qubits -- which are the base of most quantum computers architectures~\cite{Dwave,Ibm,Google} -- they do not suffer from the above mentioned drawbacks present in QWs: interactions in QCA are local and translation invariant (both in space and time). These QCA advantages, when combined with the various algorithms developed for QWs, might render the implementation of large quantum algorithms viable to current technology.

The article is organized as follows: In Section~\ref{sec:puqca} we present the QCA formal definition and work with some examples. Moving forward in Section~\ref{sec:translation} we show how to translate the QWs formalism to the QCA. We begin with the coined model, starting with a concrete example in the one dimensional lattice, establishing the motion equation in both quantum models of computation to see that they indeed give to us the same dynamics time by time. Then, afterwards we move to the general recipe. Following the same previous steps we do the same for the staggered quantum walk with Hamiltonian (SQWH). Finally in Section~\ref{sec:conc} we state some final considerations with possibles advantages and applications for these translations.

\section{Quantum cellular automata: definition and examples}\label{sec:puqca}
In classical cellular automata (think for instance on the Wolfram's elementary CA~\cite{Wolfran}), the evolution of the cell at position $x$ is given in two steps: i) First the state of the neighboring cells of $x$ is read. ii) Second, conditioned on this first read measurement, the state of the cell $x$ is updated according to a fixed local rule.  This is then repeated for all the cells in the automata in a translation invariant way, i.e., using the same updating rule for all cells and without a specific updating ordering. The locality of the automata is guaranteed by the neighboring scheme $\mc{N}$ which states that only the cells in the set $\mc{N}_x = \mc{N}+x$ will influence the update of cell at position $x$.

This idea of cellular automata clearly can not be directly translated to the quantum realm: the state of quantum systems cannot be read without disturbing the system's state. Moreover, the possible non-commutativity of quantum operations imposes some constraints on the translation invariant update. Ultimately, the challenges to properly define a quantum cellular automata  are rooted at the impossibility of cloning quantum states~\cite{no_cloning}.  Besides that, the classical automata dynamics may be irreversible, which is in contrast with a unitary quantum evolution. One way to define reversible automata is by employing the Margolus block scheme~\cite{Toffoly}. Here we define a Partitioned Unitary Quantum Cellular Automata (PUQCA)  combining the Margolus block scheme, the unitary aspects of the QCA introduced by Pérez-Delgado and Donny Cheung~\cite{QCAPerez}, and some dynamical attributes present in the QCA initiated by Watrous~\cite{Watrous}. This combination makes clear the locality and translation invariance properties, besides  giving a direct quantum generalization of classical partitioned cellular automata.

Two features will become central to our definition of partitioned quantum cellular automata: First each cell is to be  divided into $n$ subcells. This subdivision of each cell generates a finer description for the space where the automaton is defined -- be it a lattice or the more general case of a graph. Second,  we will employ different tilings over this finer description of the space. Each tiling covers the full space,  with an individual tile covering a finite subset of the subcells. The tiling is then a uniform partition of the set of subcells, with each tile an element of this partition.

With these elements, we are in position to define our partitioned unitary quantum cellular automata.
\begin{definition}[PUQCA] A Partitioned Unitary Quantum Cellular Automata  is a 5-tuple $\left(L,\mathcal{N},\Sigma,\left\{ \mc{T}_{i}\right\} ,\left\{ W_{i}\right\} \right)$ consisting of:
	\begin{enumerate}
		
		\item a $d$-dimensional lattice of cells indexed by integers $L=\mathbb{Z}^{d}$;
		
		\item a finite neighborhood scheme $\mathcal{N}\subseteq L$;
		
		\item a finite set $\Sigma$ of orthogonal basis states with $\mc{H}_\Sigma=\textnormal{span}\{\ket{\sigma}_{\sigma\in\Sigma}\}$. Each cell is divided in $n$ subcells, and to the $i$-th subcell we assign a copy $\mc{H}_{\Sigma_i}$ of $\mc{H}_{\Sigma}$. The total space associated to each cell is then $\mc{H}_\Xi=\bigotimes_{i\in\{0,\ldots,n-1\}}\mc{H}_{\Sigma_i}$;
		
		\item a finite set of tilings $\left\{ \mc{T}_{i}\right\}_{i=0}^{N-1} $. Each tiling is the union of identical non-overlapping tiles, $\mc{T}_i=\bigcup_j T_j^{(i)}$, with each tile  $T_j^{(i)}$ containing only subcells of neighboring cells.
		
		\item a set of local unitary functions $\left\{ W_{i}\right\}_{i=0}^{N-1}$. The same unitary $W_{i}$ is applied to each tile $T_j^{(i)}$  of the tiling $\mc{T}_i$;   
	\end{enumerate}
\end{definition}
With this definition, the transition function $\mathcal{E}:\left(\mathcal{H}_{\Xi}\right)^{\otimes L}\mapsto \left(\mathcal{H}_{\Xi}\right)^{\otimes L}$,    which updates the automaton state from the time $t$ to $t+1$, is the given by 
\beq
\mathcal{E}=\prod_{i=0}^{N-1}\left(\bigotimes_{T_{j}^{(i)}\in \mc{T}_i}W_{i}\right).
\eeq

\begin{figure}[ht]
	\noindent \centering{}\includegraphics[scale=0.7]{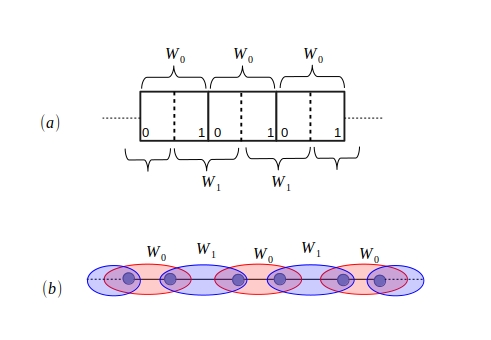}\caption{\small\label{fig:LvsG}\textbf{1-dimensional automaton.} In top-panel ($a$) it is shown how each cell is split in two subcells, and how the unitary operators $W_{i}$ are applied in accordance with the two tilings. In the bottom-panel ($b$) it is shown the same 1-d automaton, but now in the graph perspective. The tiling $\mc{T}_0$ is represented by the red ellipses, while the tiling $\mc{T}_1$ is shown in blue.}
\end{figure}

To work with these tilings more precisely, it is convenient put labels in each subcell.  Given the cell at position $i\in L$, its subcells are denoted by $i_j$, with $j\in \{0,\ldots,n-1\}$.  For instance, suppose we have an one-dimensional lattice where each cell has two subcells, and the neighbor scheme is $\mathcal{N}_{i}=\left\{i-1,i,i+1\right\}$. In this case two tilings are sufficient to evolve the automaton: The first tiling is given by $\mc{T}_0=\bigcup_{i\in \Ir} T_i^{(0)}$ with each tile defined as $T_i^{(0)}=\{i_0,i_1\}$. For the second tiling, $\mc{T}_1=\bigcup_{i\in \Ir} T_i^{(1)}$, each tile is given by $T_i^{(1)}=\{i_1,(i+1)_0\}$. It is then clear that the first tiling is responsible for ``reading'' the state of each cell, while the second one is responsible for the interaction between the neighboring cells -- resembling the unitary quantum cellular automata defined in~\cite{QCAPerez}. Now that the tilings' structure is established, the action of the unitary functions is apparent:
\begin{align*}
W_{0}:&\left(\mathcal{H}_{\Sigma_{0}}\right)_{i}\otimes\left(\mathcal{H}_{\Sigma_{1}}\right)_{i}&\rightarrow&\left(\mathcal{H}_{\Sigma_{0}}\right)_{i}\otimes\left(\mathcal{H}_{\Sigma_{1}}\right)_{i}\\
W_{1}:&\left(\mathcal{H}_{\Sigma_{1}}\right)_{i}\otimes\left(\mathcal{H}_{\Sigma_{0}}\right)_{i+1}&\rightarrow&\left(\mathcal{H}_{\Sigma_{1}}\right)_{i}\otimes\left(\mathcal{H}_{\Sigma_{0}}\right)_{i+1},
\end{align*} 
for all $i\in \Ir$. Therefore, in this example we can explicit our transition function as 
\begin{equation}
\label{Trans_1d}
\mathbb{\mathcal{E}}=\left(\bigotimes_{T_{i}^{(1)}\in \mc{T}_1}W_{1}\right)\left(\bigotimes_{T_{i}^{(0)}\in \mc{T}_0}W_{0}\right).
\end{equation}
All this construction can be apprehended from Fig.(\ref{fig:LvsG}-$a$).

\begin{figure}[ht]
	\noindent \centering{}\includegraphics[scale=0.7]{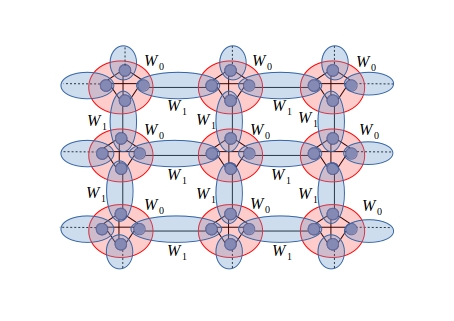}\caption{\small\label{fig:two_G}\textbf{Graph perspective: 2-dimensional automaton.} For the automaton defined over the 2-dimensional lattice, with four neighbors, each cell is transformed into a complete graph with 4 vertices (K4). The first tiling is shown in red, with corresponding unitary operation $W_0$ acting only on the subcells of each cell. The second tiling is depicted in blue, with the unitary operation $W_1$ being responsible for the interaction between neighboring cells.}
\end{figure}

The above definition of the PUQCA immediately generalizes to  QCA over a regular graph $G=G(V,E)$. In this situation, the neighborhood scheme is represented by the edge set $E$, and the tilings are defined over partitions of the graph $G$ into complete subgraphs -- in every tiling all the vertices are included, and the union of the tilings must contain all the edges. Within graph theory, tilings are usually called tessellations (see the definitions used~\ref{sub_SQWH} below). We also can see how the one-dimensional example is represented by the ``graph perspective'' in Fig.(\ref{fig:LvsG}-b). Another example of a QCA in the graph picture is shown in Fig.~\eqref{fig:two_G}.

In what follows we will mostly employ the graph perspective, as it makes clear the translation procedures from quantum walks to quantum cellular automata.

\section{Translating Quantum Walks into Quantum Cellular Automata}\label{sec:translation}

This section contains our main result: how to translate a large class of quantum walk models into quantum cellular automata models. We focus on the Coined Quantum Walk over a $d$-regular graph (CQW$_\text{d}$) and on the Staggered Quantum Walk with Hamiltonians (SQWH), as these two models were already shown to encompass various models of quantum walks~\cite{SQW_H}. The amount of resources in the quantum walk is compared to the correspondent quantum cellular automaton.

\subsection{$\text{CQW}_\text{d}  \subseteq  \text{QCA}$}

\subsubsection{Coined quantum walk over a $d$-regular graph.}

Coined quantum walks were the first to be devised~\cite{Coined}, likely due to their similarity to the classical case. The main difference from the classical to the quantum walk lies in the possibility, in the latter, of superpositions of the coin values and walker positions.  Since these early times, the definition of quantum walks evolved, specially to cope with general dynamics in arbitrary finite graphs~\cite{Philipp2016}. Given a graph $G=G(V,E)$ its vertices are associated to the walker's classical positions, while its edges define the possible directions the walker can take in a single step. To the position $i\in V$ and direction $(i,j)\in E$ we associate the state $\ket{(i,j),i}$. Defining $\mc{N}^G_i$ as the graph-neighborhood of a vertex $i$, $\mc{N}^G_i=\{j|(i,j)\in E\}$, the total Hilbert  space associated to the quantum walk is given by 
\[\mc{H}_G =\vspan(\{ \ket{(i,j),i} | (i,j)\in E,\; i\in V,\; j\in\mc{N}^G_i\}). \]
As each edge is connected to exactly two vertices, the dimension of $\mc{H}_G$ is $2|E|$\footnote{Note that the space $\mc{H}_G$ is not given by the tensor product of a space associated to the edges and another space associated to the vertices.}.

The dynamics of a quantum walker is given by first the ``flip'' of a quantum coin, followed by a coin-dependent coherent displacement. Given a walker in the state $\ket{(i,j),i}$, the coin operator must  create a superposition of all allowed directions, i.e., a superposition of the states $\ket{(i,k),i}$ for all $k\in \mc{N}_i^G$. The action of the coin operator is more easily understood if we decompose $\mc{H}_G$ in blocks related to each vertex: $\mc{H}_G=\oplus_{i\in V}\mc{H}_G^{(i)}$, with $\mc{H}_G^{(i)} =\vspan(\{ \ket{(i,j),i} | (i,j)\in E,\; j\in\mc{N}^G_i\})$. For a $d$-regular graph, each such a block is a $d$-dimensional Hilbert space. The action of the coin operator $C:\mc{H}_G\rightarrow\mc{H}_G$ can be written as:
\beq
C=\bigoplus_{i\in V}C_{i},
\eeq
where each $C_{i}$ is a unitary acting on $\mc{H}_G^{(i)}$. If the coin flip is independent of the vertex, which is usually the case, we take the action of the coin operator to be identical in each vertex-subspace. To finish the walker step we must define the  shift operator $S_\pi:\mc{H}_G\rightarrow\mc{H}_G$. As the walker can only walk ``over'' the edges, the action of the most general shift operator is given by:
\beq
S_\pi\ket{(i,j),i} = \ket{(\pi(i),j),j}, \forall i\in V \text{ and } \forall (i,j)\in E.
\eeq
Here $\pi$ is a permutation that fixes the walker's direction after the step, i.e., $\pi$ is a permutation between the $d$ neighbors of the receiving vertex such that $\pi(i)\in \mc{N}^G_j$. As the permutation $\pi$ fully determines the shift operator, in a $d$-regular graph there are only $d!$ different shift operators. Moreover, since this permutation does not change the vertex value, we can choose the identity permutation and absorb some possible relabeling in the coin operator. The shift for which we take the identity permutation, $S_{I}\ket{(i,j),i} = \ket{(i,j),j}$, is known as the flip-flop, given that $S_{I}^2=\idty$. Given all that, the one time step evolution of the quantum walker is given by the unitary $U_G:\mc{H}_G\rightarrow\mc{H}_G$, whose action can be written as:
\beq
U_G= S_I\cdot C.
\eeq

In what follows we show how to translate each element of the coined quantum walk into a quantum cellular automata description. We start with a simple example, followed by a general translation procedure.

\subsubsection{One dimensional example.}

The most paradigmatic example of coined quantum walk is the one-dimensional case, with a two-states coin. This construction closely resembles the classical random walk, and it was the main inspiration for the generalization to the quantum domain~\cite{Coined}. In this instance the walk happens in a one-dimensional lattice, which can be seen as a infinite 2-regular graph. In such a graph perspective, the set of vertices is $V=\Ir$, and the set of edges is $E=\{(i,i+1)|i\in V\}$, leading to a vertex-neighborhood $\mc{N}^G_i=\{i-1,i+1\}$. Moreover, if  the walker is on the vertex $i$ moving to the right we associate to it the state $\ket{(i,i+1),i}$, while if it is moving to the left we describe it by the state $\ket{(i,i-1),i}$. For the walker in such a position, the coin operator must superpose the two directions of movement, $\ket{(i,i+1),i}$ and $\ket{(i,i-1),i}$. This can be done by choosing $C_i: \mc{H}_G^{(i)}\rightarrow \mc{H}_G^{(i)}$ as a SU(2) operator
\[
C_i= \begin{pmatrix}
q & p\\
p & q
\end{pmatrix}\; \forall i\in V,
\]
with $p,\; q \in \Cx$ respecting the unitarity constraints $\left|p\right|^{2}+\left|q\right|^{2}=1$ and $p^*q+q^*p=0$. The total coin operator is then $C=\bigoplus_{i\in V}C_i$. Traditionally, the shift operator does not chance the coin value, i.e., it keeps the direction of the movement. This shift is known as the ``moving'' operator, and it acts on $\mc{H}_G$ as follows:
\begin{eqnarray}
S_X\ket{(i,i-1),i}&=&\ket{(i-2,i-1),i-1},\\
S_X\ket{(i,i+1),i}&=&\ket{(i+2,i+1),i+1}.\nonumber
\end{eqnarray}
The  relation of between the moving shift $S_X$ to the flip-flop $S_I$ one is simple: $S_X = X\cdot S_I$, where $X=\bigoplus_{i\in V}X_i$ with
\[
X_i= \begin{pmatrix}
0 & 1\\
1 & 0
\end{pmatrix}\; \forall i\in V.
\]
A single step evolution is then given by $U=X\cdot S_I \cdot C$~\footnote{Note that since the evolution for $t$ time steps is given by $U^t$, then the operator $X$ can be absorbed in the coin operator by suitably applying corrections on the initial and final state: $U^t =X \cdot (S_I\cdot CX)^t X^{-1}$.}. If the state of the system at time $t$ is  
\begin{eqnarray*}
\left|\psi(t)\right\rangle &=&\sum_{i\in V}\left(\psi_{\left(i,i-1\right)}\left(i,t\right)\ket{(i,i-1),i}\right.\\&+&\left.\psi_{\left(i,i+1\right)}\left(i,t\right)\ket{(i,i+1),i}\right),
\end{eqnarray*}
where $\psi_{(i,j)}(i,t):=\bra{(i,j),i}\psi (t)\>$ is the amplitude of the walker being located at the vertex $i$ pointing to the vertex $j$ at time $t$. Then the state at $t+1$ is obtained by $U\ket{\psi(t)}$, which gives
{\footnotesize
\begin{align*}
&\sum_{i\in V}\left[\left(q\psi_{\left(i,i-1\right)}\left(i,t\right)+p\psi_{\left(i,i+1\right)}\left(i,t\right)\right)\ket{(i-2,i-1),i-1}\right.\\
&+\left.\left(p\psi_{\left(i,i-1\right)}\left(i,t\right) + q\psi_{\left(i,i+1\right)}\left(i,t\right)\right)\ket{(i+2,i+1),i+1}\right].
\end{align*}}
It is now immediate to obtain the recurrence relations that govern the quantum walk dynamics:
{\footnotesize
\begin{align*}
	\psi_{(i-2,i-1)}\left(i-1,t+1\right)=&q\psi_{(i,i-1)}\left(i,t\right)+p\psi_{(i,i+1)}\left(i,t\right),\\
	\psi_{(i+2,i+1)}\left(i+1,t+1\right)=&p\psi_{(i,i-1)}\left(i,t\right)+q\psi_{(i,i+1)}\left(i,t\right).
\end{align*}}
 These are the dynamical equations that fully describe the walker dynamics in one dimensional lattice with a moving shift.

Now we want to reproduce the dynamics achieved above within the cellular automata formalism. In order to do that,  each vertex $i\in V$ we change to a cellular automaton cell. As each vertex has two neighbors (leading to a two-dimensional coin), each automaton cell is divided into two sub-cells. A cell $i$ is then composed by two subcells labeled by  $(i)_{(i-1)}$ and $(i)_{(i+1)}$. In each subcell we place a two-dimensional quantum system (a qubit), i.e., $\mc{H}_{(i)_{(i-1)}}\cong\mc{H}_{(i)_{(i+1)}}\cong\Cx^2$, for all $i\in V$. We then encode the quantum walk state $\ket{(i,i-1),i}$ with the automaton state $\ket{\ldots,(0,0)_{i-1},(1_{i-1},0_{i+1})_i,(0,0)_{i+1},\ldots}$, while the state $\ket{(i,i+1),i}$ is written as $\ket{\ldots,(0,0)_{i-1},(0_{i-1},1_{i+1})_i,(0,0)_{i+1},\ldots}$.  In this way, an excited subcell (a qubit in the 1 state) indicates the walker position and its movement direction. All the encoding is done within the single excitation subspace.

For the dynamics translation, for clarity, we will employ one tiling for each quantum walk unitary. As such, we require three tilings.  The first one, $\mc{T}_0$, is related to the coin operator. As this only changes the edge state, without moving the walker, the tiles are the very subcells of each cell: $T_i^{(0)}=\{(i)_{i-1},(i)_{i+1}\}$. The unitary $W_0$ associated to this tiling is of the same form as the unitary $C_i$ employed in the coin $C$:
\[
W_{0}=\begin{pmatrix}1 & 0 & 0 & 0\\
0 & q & p & 0\\
0 & p & q & 0\\
0 & 0 & 0 & 1
\end{pmatrix},
\]
also fulfilling the requirements that $\left|p\right|^{2}+\left|q\right|^{2}=1$ and $p^*q+q^*p=0$. Note that now such operator acts on a two-qubit system, $W_0:\mc{H}_{\Sigma_{(i,i-1)}}\otimes\mc{H}_{\Sigma_{(i,i+1)}}\rightarrow\mc{H}_{\Sigma_{(i,i-1)}}\otimes\mc{H}_{\Sigma_{(i,i+1)}}$, and it is written in the basis $\{\ket{(0,0)_i},\ket{(0,1)_i},\ket{(1,0)_i},\ket{(1,1)_i}\}$. Moreover, notice that $W_0$ only acts non-trivially on the single excitation subspace.  Writing a general state $\ket{\Psi(t)}$ for the automaton at time $t$ as:
{\footnotesize
\begin{align*}
\sum_{i\in\mathbb{Z}}&\left[\Psi_{(i,i-1)}\left(i,t\right)\left|\ldots,\left(0,0\right)_{i-1},\left(1_{i-1},0_{i+1}\right)_{i},\left(0,0\right)_{i+1},\ldots\right\rangle +\right.\\
&+\left.\Psi_{(i,i+1)}\left(i,t\right)\left|\ldots, \left(0,0\right)_{i-1},\left(0_{i-1},1_{i+1}\right)_{i},\left(0,0\right)_{i+1},\ldots\right\rangle \right],
\end{align*}}
where $\Psi_{(i,i\pm1)}\left(i,t\right)$ is the probability amplitude of finding an ``excitation'' at the subcell  $i\pm1$  of the $i$-th cell at time $t$, after the evolution corresponding to the first tiling, $\left(\bigotimes_{T_{j}^{(0)}\in \mc{T}_0}W_{0}\right)\ket{\Psi(t)}$,  we get:
{\footnotesize
\begin{align*}
\sum_{i\in\mathbb{Z}}&\left[\Psi_{(i,i-1)}\left(i,t\right)\left|\ldots,\left(0,0\right)_{i-1},\left(q_{i-1},p_{i+1}\right)_{i},\left(0,0\right)_{i+1},\ldots\right\rangle +\right.\\
&+\left.\Psi_{(i,i+1)}\left(i,t\right)\left|\ldots, \left(0,0\right)_{i-1},\left(p_{i-1},q_{i+1}\right)_{i},\left(0,0\right)_{i-1},\ldots\right\rangle \right].
\end{align*}}
Now we use a second tiling $\mc{T}_1$, to simulate the flip-flop shift operator $S_I$. The second tiling is thus responsible for the interaction between the cells. As such,  the associated unitary operation $W_1$ must act on tiles that contain subcells from neighboring cells. As the walker wave-function can only spread by one vertex in a single step, to the left and to the right, the automaton neighborhood is given by $\mc{N}=\{j-1,j,j+1\}$. From the definition of the flip-flop shift operator $S_I$  and the encoding of the coin-walker state in the automaton language, we note that the probability amplitude  in the subcell $i+1$ of the $i$-th cell ($i_{i+1}$) is to be  transfered to the  subcell $i$ of the $(i+1)$-th cell ($(i+1)_{i}$), and vice-versa -- in the graph perspective, we create an edge $(i, i+1)$ for all $V$. Given that, for the tiling $\mc{T}_1$ we set the tiles $T_i^{(1)}=\{(i)_{i+1},(i+1)_{i}\}$, and the unitary function $W_1$ is simply the {\sc swap} operator. After the evolution due to these two tilings, $\left(\bigotimes_{T_{j}^{(1)}\in \mc{T}_1}W_{1}\right)\left(\bigotimes_{T_{j}^{(0)}\in \mc{T}_0}W_{0}\right)$, a general state $\ket{\Psi(t)}$ the state of the automaton reads:
{\footnotesize
\begin{align*}
\sum_{i\in V}&\left[\Psi_{(i,i-1)}\left(i,t\right)\left|\ldots,\left(0_{i-2},q_{i}\right)_{i-1},\left(0,0\right)_{i},\left(p_{i},0_{i+2}\right)_{i+1},\ldots\right\rangle +\right.\\
&+\left.\Psi_{(i,i+1)}\left(i,t\right)\left|\ldots, \left(0_{i-2},p_i\right)_{i-1},\left(0,0\right)_{i},\left(q_{i},0_{i+2}\right)_{i+1},\ldots\right\rangle \right].
\end{align*}}
Lastly, we need to simulate the action of the $X$ operator with a third tiling, $\mc{T}_2$. Like for the coin operator, the $X$ operator does not chance the walker's position. As such, its corresponding operator acts only on individual cells: $T_i^{(2)}=\{(i)_{i-1},(i)_{i+1}\}$. Moreover, as the action of each $X_i$ is simply to change the movement direction, here we only need to take $W_2$ as a {\sc swap} operator acting in each cell to get:
{\footnotesize
	\begin{align*}
	\sum_{i\in V}&\left[\Psi_{(i,i-1)}\left(i,t\right)\left|\ldots,\left(q_{i-2},0_{i}\right)_{i-1},\left(0,0\right)_{i},\left(0_{i},p_{i+2}\right)_{i+1},\ldots\right\rangle +\right.\\
	&+\left.\Psi_{(i,i+1)}\left(i,t\right)\left|\ldots, \left(p_{i-2},0_i\right)_{i-1},\left(0,0\right)_{i},\left(0_{i},q_{i+2}\right)_{i+1},\ldots\right\rangle \right].
	\end{align*}}
That is the automaton state after on time step, i.e., after the action of $\mc{E} =\left(\bigotimes_{T_{j}^{(2)}\in \mc{T}_2}W_{2}\right)\left(\bigotimes_{T_{j}^{(1)}\in \mc{T}_1}W_{1}\right)\left(\bigotimes_{T_{j}^{(0)}\in \mc{T}_0}W_{0}\right)$. It is now immediate to obtain the recurrence relations that describe the automaton dynamics:
{\footnotesize
	\begin{align*}
	\Psi_{(i-2,i-1)}\left(i-1,t+1\right)=&q\Psi_{(i,i-1)}\left(i,t\right)+p\Psi_{(i,i+1)}\left(i,t\right),\\
	\Psi_{(i+2,i+1)}\left(i+1,t+1\right)=&p\Psi_{(i,i-1)}\left(i,t\right)+q\Psi_{(i,i+1)}\left(i,t\right).
	\end{align*}}
These are exactly the same recurrence relations for the 1-d coined quantum walk. Both dynamics are thus identical at every time-step.


\subsubsection{General Recipe}

Now we give a prescription to find the  QCA correspondent to a given coined QW on a $d$-regular graph. As input we take a $d$-regular graph $G=G(E,V)$ (or lattice $L$) where the quantum walk is defined; the coin operator $C$; and the shift operator $S_\pi$. As output we must return a complete PUQCA whose evolution is the same as the  CQW. The steps to find this translation are enumerated below.

\begin{enumerate}
	\item The number of cells in the automaton is given the number of vertices, $|V|$, of the graph $G$. As the graph is $d$-regular, each cell is split in $d$ subcells.  We place one qubit in each subcell, and then a total of $|V|.d$ qubits are employed (see the resources discussion below).

	\item The automaton neighborhood scheme $\mc{N}$ is determined by the graph-neighborhood $\mc{N}^G$, by the simple inclusion of the ``central'' cell: $\mc{N}_i=\mc{N}^G_i\cup \{i\}$.

	\item To each CQW$_d$ state $\ket{(i,j),i}$ we associate the automaton (single excitation subspace) state $\ket{\ldots (0_h,\ldots,1_j,\ldots,0_{m})_i\ldots}$, where $h,m$ and all other subindex labeling the subcells of cell  $i$ belong to $\mathcal{N}^G_i$. While the subindex $i$ gives us in which cell the excitation is located, the subindex $j$ tells us its subcell location (corresponding to the movement direction). In this way the space associated to each cell is $\mc{H}_\Xi=\left(\mathbb{C}^{2}\right)^{\otimes d}$.
	
	\item To simulate the CQW$_d$ dynamics within the automata language three tilings are sufficient~\footnote{Here again the action of the third tiling can be absorbed in the action of the first one, plus modifications in the initial and final state. We however present the translation with three tilings for clarity reasons.}. The first tiling is related to the action of the coin, with each tile given by all subcells that belongs to the same cell: $T_{i}^{(0)}=\{ (i)_{j} | j\in \mc{N}^G_i\}$. The second tiling is devoted to the simulation of the flip-flop shift acting on neighboring cells. From the definition of $S_I$ we can see that in terms of QCA we have the subcell $j$ in the cell $i$ interacting with the subcell $i$ of the $j$-th cell, and vice-versa. Therefore $T_{(i,j)}^{(1)}=\{(i)_{j}, (j)_{i}\} $ where  $(i,j)\in E$. The third tiling is responsible for simulating the permutation that connects $S_I$ to $S_\pi$. As this operation is ``local'' in each vertex, then $T_{i}^{(1)}=\{ (i)_{j} | j\in \mc{N}^G_i\}$.

	\item To each tiling we associate one unitary operator. To the first tiling, the unitary operator $W_0:\mc{H}_\Xi\rightarrow \mc{H}_\Xi$ is directly related to the unitary operator $C_i$ by  employing the unary representation for the cell states (see two steps above) within the single excitation subspace. Out of this subspace the action of $W_0$ is trivial, being completed by ones in the diagonal entries. Since the flip-flop operator is translated as a swap between the two subcells of neighboring cells, the unitary $W_1:\mc{H}_{(i)_{j}}\otimes \mc{H}_{(j)_{i}}\rightarrow\mc{H}_{(i)_{j}}\otimes \mc{H}_{(j)_{i}}$, for $(i,j)\in E$, is the {\sc swap} gate between them. Lastly, $W_2:\mc{H}_\Xi\rightarrow \mc{H}_\Xi$ implements the permutation $\pi$ on the cell $i$, by encoding the operator $\pi_i$ in the same way as for the coin operator.
	 
\end{enumerate}
These steps give the full translation between a CQW$_d$ and a QCA.

\bigskip

Before we move on, we compare the resources required in each model. The dimension of the Hilbert space in the quantum walk model is $2 |E|$, which is equal $|V| d$ due to the assumed $d$-regularity of the graph. For the corresponding QCA at first sight we would need a Hilbert space of dimension $|V|.2^d$. Nevertheless, as we only need the single excitation subspace for our construction, whose dimension is $d$, we in fact only use an effective  Hilbert space  with dimension $|V|.d$. Therefore, both models require the same amount of resources.
\subsection{$\text{SQWH} \subseteq  \text{QCA}$}\label{sub_SQWH}

Sometime after the development of the coined quantum walk model, it was realized that similar coherent dynamics could be obtained even without a coin space. In such models the coin is removed and the walker's dispersion over a graph is obtained via the alternation of operators directly acting on the walker's position space. The first ``coinless'' quantum walk model was proposed by  Szegedy~\cite{Szegedy1},  followed by the staggered quantum walk~\cite{coinless,staggered,stag_graph,SQW_H}, and, more recently, by the staggered quantum walks with Hamiltonians (SQWH)~\cite{SQW_H}. The latter was shown to encompass the previous two models of coinless quantum walks, and even some coined models~\cite{SQW_H}. Furthermore, the SQWH has been shown to be useful for quantum search~\cite{sqwh_search}, and an experimental proposal is already in place~\cite{sqwh_exp}. Our aim here is thus to show that the PUQCA can simulate the SQWH, and with that inherit the simulation of various coinless, and coined, models.

We start by describing the SQWH over a graph $G=G(V,E)$~\cite{SQW_H}. As before, to each vertex $i\in V$ we associate a unit vector $\ket{i}$, with $\<i|j\>=\delta_{ij}$ for all $i,j\in V$. As such, to the vertices of $G$ we associate the Hilbert space $\mc{H}_V = \vspan(\{\ket{i} \;|\; i \in V \})$. Crucial to the SQWH is the  concept of a graph \textit{tessellation}:  A graph tessellation $\g{T}$ is a partition of the graph into complete subgraphs, i.e., into cliques. Such a partition directly induces a partition of $\mc{H}_V$: let $\alpha$ be an element of $\g{T}$, then $\mc{H}_V = \bigoplus_{\alpha \in \g{T}} \mc{H}_\alpha$, where $\mc{H}_\alpha= \vspan(\{\ket{i} \;|\; i\in \alpha\})$. Each element $\alpha$ is called a polygon, as it is related to a clique. It is now simple to define a rank-one projector $\proj{\alpha}$ into $\mc{H}_\alpha$, by defining the vector
\[
\ket{\alpha}=\sum_{i \in\alpha} a(i)\ket{i},
\]
where $a(i)\in\Cx$ and $\sum_{i\in\alpha}|a(i)|^{2}=1$. Note that the coefficients $a$ do not depend on the polygon in a given tessellation, but they do depend on the label given to each vertex within polygon~\cite{SQW_H}.  A dynamics that does not connect different subspaces $\mc{H}_\alpha$ can be given by the Hamiltonian operator associated with this tessellation as
\begin{equation}
\label{ham_stag}
H_{\g{T}}=2\sum_{\alpha\in \g{T}} \proj{\alpha}-\idty.
\end{equation}
Such operator is known as the orthogonal reflection of the graph~\cite{stag_graph}, and it is  Hermitian and unitary, implying that $H_{\g{T}}^2 =\idty$. The dynamics generated by this Hamiltonian is then $U_\g{T}= \exp(\ii \theta H_\g{T})$, with $\theta \in [0,2 \pi]$. This propagator respects the partition of $\mc{H}_V$ into subspaces related to the tessellation polygons, as $U_\g{T}=\bigoplus_{\alpha\in \g{T}}U_\alpha$ with 
\beq
\label{eq:Ualpha}
U_\alpha = e^{-\ii \theta}\idty_\alpha + 2 \ii \sin(\theta)\sum_{i,j\in \alpha}a^*(i)a(j)|i\>\!\<j|;
\eeq
where $\idty_\alpha:=\sum_{i \in\alpha}|i\>\!\<i|$ is the identity operator in the $\alpha$ subspace.

If the dynamics of the walker was to be given simply by the propagator $U_\g{T}$, then a walker starting in a vertex $i\in V$ would remain stuck in the polygon that contains such a vertex. As Szegedy noticed~\cite{Szegedy1}, a walker dispersion over a graph can be obtained without a coin if we alternate propagation operators, with each of them acting within a different subspace-partition of $\mc{H}_V$. At this point we observe that a given  tessellation  contains all the vertices of a graph, but not necessarily all its edges. In~\cite{SQW_H} it was defined a set of tessellations, a \textit{tessellation cover} $\{\g{T}_0,...,\g{T}_{N-1}\}$,   whose union also covers  the edge set. Each tessellation $\g{T}_k$ induces a different subspace-partition of $\mc{H}_V$, with associated Hamiltonian $H_{\g{T}_k}$ constructed in the same way as in Eq.\eqref{ham_stag}. One time-step evolution of the SQWH is then generated by the operator
\begin{equation}
U=\prod_{k=0}^{N-1}e^{i\theta_{k}H_{\g{T}_{k}}}
\end{equation}
with $\theta_{k} \in [0,2\pi]$ for all $k\in\{0,\ldots,N-1\}$. 

\bigskip
We are now ready to show how to translate the SQWH model into a QCA one. As previously, before giving a general recipe we first show a simple example of such translation.

\subsubsection{One dimensional example}

\begin{figure}
	\includegraphics[scale=0.4]{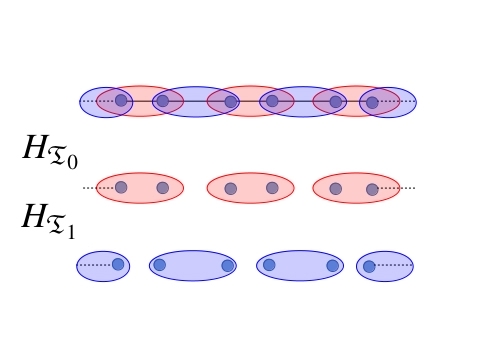} \caption{\label{One_SQWH} {\small This picture shows the two tessellations used in the $1-d$ example of translation between the SQWH and the PUQCA. Within the SQWH the operator $U_0$ ($U_1$) acts on the red (blue) polygons. In the PUQCA picture, this is translated to the action of $W_0$ ($W_1$) on the qubits in the red (blue) tiles.}}
\end{figure}

In this example, we consider a SQWH over a 1-d lattice. The vertex set is $V=\Ir$, and thus $\mc{H}_V = \vspan(\{\ket{i} \;|\; i\in\Ir\})$.  The smallest tessellation cover for such 2-regular (infinite) graph is composed of two tessellations $\{\g{T}_0,\g{T}_1\}$, with $\g{T}_0=\{\includegraphics[scale=0.3,valign=c]{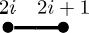}\;|\; i\in \Ir\}$ and $\g{T}_1=\{\includegraphics[scale=0.3,valign=c]{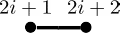}\;|\; i\in \Ir\}$. These tessellations are shown Fig.(\ref{One_SQWH}). Each tessellation induces a different partition of $\mc{H}_V$ as $\bigoplus_{\alpha_k \in \g{T}_k} \mc{H}_{\alpha_k}$, with $k\in\{0,1\}$. For this example we take general projectors  into each polygon-subspace via the vectors
\begin{align*}
&\ket{\includegraphics[scale=0.3,valign=c]{alpha0}} = a_0\ket{2i}+\tilde{a}_0\ket{2i+1},\\
&\ket{\includegraphics[scale=0.3,valign=c]{alpha1}} = a_1\ket{2i+1}+\tilde{a}_1\ket{2i+2},
\end{align*}
for all $i\in \Ir$, and where the coefficients are constrained to $|a_k|^2+|\tilde{a}_k|^2=1$ with $k\in\{0,1\}$. With these projectors we follow Eq.\eqref{ham_stag} to construct the evolution operator as  
\[
U=e^{\ii\theta_1 H_{\g{T}_1}}e^{\ii\theta_0 H_{\g{T}_0}}.
\]
As the evolution propagator is composed by the product of similar operators, $e^{\ii\theta_k H_{\g{T}_k}}$, acting in similar ways in different partitions of $\mc{H}_V$, below we only show how to translate a single of these operators, say $e^{\ii\theta_0 H_{\g{T}_0}}$,  into automata language. Still within the SQWH language, with the aid of Eq.\eqref{eq:Ualpha}, the propagator in each subspace is
\begin{align*}
U_{\includegraphics[scale=0.2,valign=c]{alpha0}}=& e^{-\ii \theta_0}\Big(\proj{2i}+\proj{2i+1}\Big) + \\
&2 \ii \sin(\theta_0)\Big(|a_0|^2 \proj{2i} +a_0^*\tilde{a}_0|2i\>\!\<2i+1|+\\
&a_0\tilde{a}^*_0|2i+1\>\!\<2i|+|\tilde{a}_0|^2 \proj{2i+1}\Big).
\end{align*}
The evolution operator for this tessellation is then $U_{\g{T}_0}=\bigoplus_{i \in \Ir} U_{\includegraphics[scale=0.2,valign=b]{alpha0}}$. Let a general walker state at time $t$ be expressed as $\ket{\psi(t)}=\sum_{i\in \Ir} \psi(i,t)\ket{i}$, where $\psi(i,t)$ is the probability amplitude for vertex $i$ at time $t$. The state after the action of $U_{\g{T}_0}$ is then given by
\begin{align}
\label{eq:SQWH_1Devol}
U_{\g{T}_0}\ket{\psi(t)} =\sum_{i \in \Ir}&\Big\{\Big. \big[\big.(e^{-\ii \theta_0}+2 \ii \sin(\theta_0)|a_0|^2)\psi(2i,t)+\nonumber\\
&+2 \ii \sin(\theta_0) a_0\tilde{a}_0^*\psi(2i+1,t)\big]\big.\ket{2i}+\nonumber\\
&\big[\big.(e^{-\ii \theta_0}+2 \ii \sin(\theta_0)|\tilde{a}_0|^2)\psi(2i+1,t)+\nonumber\\
&+2 \ii \sin(\theta_0) \tilde{a}_0 a_0^*\psi(2i,t)\big]\big.\ket{2i+1}\Big\}\Big..
\end{align}
This is the evolution that we want to simulate within the quantum cellular automata model.

For the QCA simulation, in each lattice vertex we place one qubit. The two tessellations needed for the walker dynamics will now give us two tilings. The polygons of the first tessellation determine now the tiles of the first tiling: $T_{i}^{(0)}=\left\{ {2i},{2i+1}\right\}$. Similarly, for the second tiling the correspondence implies the tiles $T_{i}^{(1)}=\left\{ {2i+1},{2i+2}\right\}$. There are two possibilities to define the cellular automaton cell: first is to take the tiles of the first tiling as forming a single cell with two subcells; second is to consider each vertex as a single cell with no subcell division. Both cases are equivalent, and we take the second choice as it simplifies the simulation description. The encoding of the walker state into the QCA framework is then simply given by $\ket{i}\rightarrow \ket{\ldots,0_{i-1},1_{i},0_{i+1},\ldots}$, for all $i\in V$. A general state for the automaton at time $t$ is then written as $\ket{\Psi(t)}=\sum_{i\in \Ir}\Psi(i,t)\ket{\ldots,0_{i-1},1_{i},0_{i+1},\ldots},$ with $\Psi(i,t)$ the amplitude of finding on ``excitation'' at the $i$-th qubit at time $t$. Now we need to simulate the evolution operator $U_{\g{T}_0}$ with the action of the first tiling. To that we note that each polygon in $\g{T}_0$ corresponds exactly to a tile in $\mc{T}_0$. Therefore, the propagator $U_{\g{T}_0}=\bigoplus_{\alpha_0\in \g{T}_0}U_{\alpha_0}$ is then translated into $\bigotimes_{T^{(0)}_i\in\mc{T}_0}W_0$, where 
{\small
	\[
	W_{0}=\begin{pmatrix}1 & 0 & 0 & 0\\
	0 & e^{-i\theta_0}+2\ii\sin(\theta_0)|a_{0}|^{2} & 2\ii\sin(\theta_0) a_{0}\tilde{a}_0^* & 0\\
	0 & 2\ii\sin(\theta_0) a_{0}^*\tilde{a}_0 & e^{-i\theta_0}+2\ii\sin(\theta_0)|\tilde{a}_{0}|^{2} & 0\\
	0 & 0 & 0 & 1
	\end{pmatrix},
	\]
}
when written in the computational basis $\{\ket{0_{2i},0_{2i+1}}, \ket{0_{2i},1_{2i+1}}, \ket{1_{2i},0_{2i+1}}, \ket{1_{2i},1_{2i+1}}\}$. With such a encoding, the evolution given by $\bigotimes_{T^{(0)}_i\in\mc{T}_0}W_0$ leads a general state of the automaton at time $t$ to the state:
{\small
	\begin{align}
	\label{eq:QCA_SQWH_1Devol}
	&\sum_{i \in \Ir}\Big\{\Big. \big[\big.(e^{-\ii \theta_0}+2 \ii \sin(\theta_0)|a_0|^2)\Psi(2i,t)\nonumber\\
	&+2 \ii \sin(\theta_0) a_0\tilde{a}_0^*\Psi(2i+1,t)\big]\big.\ket{\ldots,0_{2i-1},1_{2i},0_{2i+1},0_{2i+2},\ldots}+\nonumber\\
	&\big[\big.(e^{-\ii \theta_0}+2 \ii \sin(\theta_0)|\tilde{a}_0|^2)\Psi(2i+1,t)+\nonumber\\
	&+2 \ii \sin(\theta_0) \tilde{a}_0 a_0^*\Psi(2i,t)\big]\big.\ket{\ldots,0_{2i-1},0_{2i},1_{2i+1},0_{2i+2}\ldots}\Big\}\Big..
	\end{align}
}
After decoding, this state is exactly equivalent to the SQWH state shown in Eq.\eqref{eq:SQWH_1Devol}.

This example shows that the tessellations in the SQWH play the role of the tilings in the QCA, and the set of operators $\left\{W_i\right\}$ in the QCA are obtained from the polygon-subspace operators associated with the tessellations of the SQWH. 
\subsection{General Recipe}
The general procedure to translate a SQWH into a PUQCA takes as input a $d$-regular graph $G=G(E,V)$ (or lattice $L$), a tessellation cover $\{\g{T}_k\}_{k=0}^{N-1}$ with polygons $\alpha_k$ for each tessellation, a set of coefficients $\{a_k(i)\}_{i=0}^{|\alpha_k|-1}$, and a set of angles $\{\theta_k\}_{k=0}^{N-1}$. As result, the procedure outputs a well-formed  PUQCA by following the subsequent recipe.
\begin{enumerate}
	\item The PUQCA is established over the same graph $G$ employed by the SQWH.
	
	\item Since the graph is the same, the neighborhood scheme for the corresponding automaton is simply  $\mathcal{N}_i=\{j|(i,j)\in E\}\cup\{i\}$.
	
	\item No subcell structure is required for this translation ($n=0$). In each vertex we place a qubit, and thus $\mc{H}_\Xi  \cong \Cx^2$.
	To each vertex state $\ket{i}$ on the SQWH side, we associate the automaton state $\ket{\ldots,0_{i-1},1_{i},0_{i+1},\ldots}\in \mc{H}_\Xi^{\otimes |V|}$.
	
	\item  For each tessellation $\g{T}_k$ for the SQWH,  we have an equivalent tiling $\mc{T}_k$ in the PUQCA. Moreover, the vertices belonging to the polygon $\alpha_{k}$ of the $k$-th tessellation yield the elements of the tile $T_{k}^{(i)}$.
	
	\item With the input data, following the prescription of the SQWH, we  construct the evolution operator $U_{\g{T}_k}=\bigoplus_{\alpha_k\in \g{T}_k} U_{\alpha_k}$ to each tessellation. As the polygons $\alpha_k$ are all identical for a $d$-regular graph, the operators $U_{\alpha_k}$, acting in each polygon-subspace are also identical. Such operator is then directly translated to the PUQCA model as the unitary function $W_k$, via the unary encoding described in the item 3 above.
\end{enumerate}

This completes the translation from the SQWH to the PUQCA model. As we have done before, let us analyze the resources required by the PUQCA formalism for a given a SQWH dynamics. For a graph $G=G(E,V)$, the Hilbert space associated with the SQWH is a $|V|$-dimensional one. For the PUQCA we employed $|V|$ qubits, yielding a  $2^{|V|}$-dimensional total space. However, here again, we only use the single-excitation subspace, which is $|V|$-dimensional. Once more, the effective amount of resources required by the PUQCA are the same as the original model.


\section{Discussion and conclusions}\label{sec:conc}

In this work we introduced a partitioned quantum cellular automata, the PUQCA, which is at the same time  a well-formed quantum cellular automata, and also a conceptually simple and versatile construction. Such characteristics allowed us to show that various models of quantum walks can be readily translated into a PUQCA dynamics. This has the immediate advantage to employ qubits, the most common basic architecture of quantum computers. As such it is more convenient for near term experimental implementations of, for instance, quantum search algorithms.  It is important to notice that the translation from quantum walk models to quantum cellular automata proposed here requires the same amount of effective dimensions.

One question that immediately raises from our translation results (QW$\rightarrow$QCA) is whether quantum walks and quantum cellular automata are equivalent, i.e., if there is for every QCA an correspondent QW (QW$\leftarrow$QCA). Of course this depends on how strict the definitions of the models are. For instance, it is usually accepted that in a coined quantum walk, the shift operator does not create superposition of the walker's position states. When translating CWQ into PUQCA, this implied that all the interactions between cells were simply {\sc swap} gates. Therefore, if a given PUQCA has an interaction between cells other than a {\sc swap} gate, our results suggest that there is no equivalent coined quantum walk to such a PUQCA. Another interesting perspective of our work is to allow for comparing different types of resources across different models. We can compare, for example, how the coherence created within the staggered quantum walk models is translated into entanglement between qubits in the quantum cellular automata model.

To conclude, we recall that the idea of quantum computers was arguably created by Feynman as a way (possibly the only one) to simulate complex quantum systems. In his seminal article entitled ``Simulating Physics with Computers''~\cite{Feynman}, he constructs the idea of a simulator of physical systems from the idea of a cellular automata. At that time quantum cellular automata were not yet invented. From there on, various definitions of QCA were developed, but the potential of quantum cellular automata models as simulators of complex quantum systems remains largely unexplored.  We hope that our first results in this direction,  together with a simple formulation of a QCA model and its possibility of implementation with current technology, will serve as the catalyst for the development of a whole new phenomenology of simulation of quantum systems.

\section*{Acknowledgments}

We would like to thank Osvaldo J.~Farias for various discussions on the topic of quantum cellular automata.  We acknowledge financial support from
the National Institute for Science and Technology of Quantum Information (INCT-IQ/CNPq, Brazil).

\bibliographystyle{unsrt}
\bibliography{QWviaQCAv13}

\end{document}